\documentclass[sigconf, nonacm, pdfa]{acmart}

\usepackage[utf8]{inputenc}
\usepackage[UKenglish]{babel}
\usepackage[a-2b,mathxmp]{pdfx}[2018/12/22]
\hypersetup{pdfstartview=}
\usepackage{balance}
\usepackage[inkscapelatex=false]{svg}
\usepackage{algorithm}
\usepackage{algorithmic}

\newcommand\vldbdoi{10.14778/3685800.3685823}
\newcommand\vldbpages{4014 - 4024}
\newcommand\vldbvolume{17}
\newcommand\vldbissue{12}
\newcommand\vldbyear{2024}
\newcommand\vldbauthors{\authors}
\newcommand\vldbtitle{\shorttitle} 
\newcommand\vldbavailabilityurl{}
\newcommand\vldbpagestyle{plain} 

\begin{document}
\title{Large-Scale Metric Computation in Online Controlled Experiment Platform}

\author{Tao Xiong}
\affiliation{%
  \institution{Tencent Inc.}
  \city{Shenzhen}
  \country{China}
}
\email{txiong@tencent.com}








\author{Yong Wang}
\affiliation{%
  \institution{Tencent Inc.}
  \city{Shenzhen}
  \country{China}
}
\email{darwinwang@tencent.com}







\begin{abstract}
Online controlled experiment (also called A/B test or experiment) is the most important tool for decision-making at a wide range of data-driven companies like Microsoft, Google, Meta, etc. Metric computation is the core procedure for reaching a conclusion during an experiment. With the growth of experiments and metrics in an experiment platform, computing metrics efficiently at scale becomes a non-trivial challenge. This work shows how metric computation in WeChat experiment platform can be done efficiently using bit-sliced index (BSI) arithmetic. This approach has been implemented in a real world system and the performance results are presented, showing that the BSI arithmetic approach is very suitable for large-scale metric computation scenarios.
\end{abstract}

\maketitle

\pagestyle{\vldbpagestyle}
\begingroup\small\noindent\raggedright\textbf{PVLDB Reference Format:}\\
\vldbauthors. \vldbtitle. PVLDB, \vldbvolume(\vldbissue): \vldbpages, \vldbyear.\\
\href{https://doi.org/\vldbdoi}{doi:\vldbdoi}
\endgroup
\begingroup
\renewcommand\thefootnote{}\footnote{\noindent
This work is licensed under the Creative Commons BY-NC-ND 4.0 International License. Visit \url{https://creativecommons.org/licenses/by-nc-nd/4.0/} to view a copy of this license. For any use beyond those covered by this license, obtain permission by emailing \href{mailto:info@vldb.org}{info@vldb.org}. Copyright is held by the owner/author(s). Publication rights licensed to the VLDB Endowment. \\
\raggedright Proceedings of the VLDB Endowment, Vol. \vldbvolume, No. \vldbissue\ %
ISSN 2150-8097. \\
\href{https://doi.org/\vldbdoi}{doi:\vldbdoi} \\
}\addtocounter{footnote}{-1}\endgroup

\ifdefempty{\vldbavailabilityurl}{}{
\vspace{.3cm}
\begingroup\small\noindent\raggedright\textbf{PVLDB Artifact Availability:}\\
The source code, data, and/or other artifacts have been made available at \url{\vldbavailabilityurl}.
\endgroup
}

\section{Introduction}
Online controlled experiment (A/B tests) plays a central role in data-driver decision-making in the Internet industry \cite{gupta2018anatomy, kohavi2013online, tang2010overlapping, bakshy2014designing}. With the collection of massive data from users during the experiments, efficient data processing and analyzing is crucial for reaching conclusions of experiments, and metric computation is the most important part of it.

Metric computation in online controlled experiment platform is usually performed with queries that join, filter and aggregate the experiment data in a variety of ways. In WeChat experiment platform, we have tens of thousands metrics, and also tens of thousands experiment strategies running simultaneously, each experiment strategy affects tens of millions users on average, and hundreds of metrics would be computed in each experiment. As a result, the metric computation procedure in WeChat costs hundreds of petabytes network traffic and millions of CPU hours every day, becoming the most resource-consuming procedure in our experiment platform. Because the data involved is very large, computing metrics efficiently and answering related user queries quickly is a critical issue in our context. 

Metric computation in WeChat experiment platform can be considered as some kind of specialized online analytical processing (OLAP) task, it differs in the following aspects from general purpose OLAP.
\begin{itemize}
    \item Not only a single value should be computed for an experiment metric, the variance of the metric (and the covariance between metrics) should be estimated correctly for statistical inference.
    \item Experiment data in our situation can be organized into few categories, each category of data is related to WeChat-user activities and attributes. We found that this kind of data follows the Pareto principle \cite{dunford2014pareto, newman2005power} (also known as the 80-20 rule) roughly, makes them very efficient when represented by BSI.
    \item Most of the queries on the experiment data follow some fixed paradigms, which makes it possible for us to propose specialized solutions for handling these queries efficiently and quickly.
\end{itemize}

Based on these characteristics in our scenarios, this paper first introduce the representation of experiment data, we show that our representation of data using BSI can be very efficient and satisfy the needs of statistical inference. Later we introduce the computation details and conclude that most of the metric computations can be efficiently accomplished by BSI arithmetic. Then we describe our system architecture and present the performance results, shows that how efficient computation and fast interactive query can be achieved in a real world system using reasonable resources. At last, we discuss the expressive power and limitation of the proposed method.

To the best of our knowledge this is the first paper that implements a real world large-scale experiment metric computing system using BSI arithmetic. The primary contributions of this paper can be summarized as follows:
\begin{itemize}
    \item Implement a real world large-scale metric computation system using BSI arithmetic, deal with the real world problems encountered in experiment metric computing scenarios.
    \item Present the performance of our system, shows that BSI arithmetic approach is very suitable for large-scale experiment platform.
\end{itemize}

The rest of the paper is organized as follows. We present the related work in \autoref{sec:related-work}. \autoref{sec:bsi-representation} present the details of the representation of our experiment data by BSI. Algorithms for computing metrics in different scenarios are presented in \autoref{sec:metric-computation}. In \autoref{sec:arch}, we introduce our system architecture for metric computation. The performance results in \autoref{sec:perf} shows that the efficiency and low latency characteristic of our system. The expressive power and limitation are discussed in \autoref{sec:limitation}. Our conclusions are given in \autoref{sec:conclusion}.

\section{Related Work}
\label{sec:related-work}

\subsection{Roaring Bitmap}
Roaring bitmap was introduced in \cite{chambi2016better, lemire2016consistently}. It is a bitmap scheme using hybrid compression technique that uses both uncompressed bitmaps and packed arrays inside a two-level tree. The roaring bitmap is used to represent a set of 32-bit unsigned integers. A roaring bitmap is a key-value data structure where each key-value pair represents a set $S$ of 32-bit unsigned integers that share the same most significant 16 bits. The key is made of the shared 16 bits, and the value is a container storing the remaining 16 least significant bits for each member of $S$. According to the characteristics of values stored in each container, roaring bitmap uses different container structures for better compress the values. For example, roaring bitmap uses a sorted array when the number of values does not exceed 4096 in a container (a sparse container). Furthermore, roaring bitmap operations (AND, OR, ANDNOT, XOR, etc.) can be parallelized by exploiting single-instruction-multiple-data (SIMD) instructions \cite{lemire2016simd, lemire2015decoding}. It is worth mentioning that the compactness of the integers in a roaring bitmap has great performance impact on bitmap operations due to its design, the denser the bitmap is, the faster the operations are. We use roaring bitmap as a component for building BSIs of experiment data. 

\subsection{Bit-Sliced Index (BSI)}
Bit-sliced index (BSI) was introduced in \cite{o1997improved}. It is an ordered list of bitmaps, $B^s, B^{s-1}, ..., B^1, B^0$, and is used to represent the values (normally non-negative integers, and we use BSI to represent non-negative numeric values in our system) of some column $C$ of a table $T$ (although the column $C$ might be calculated values associated with rows of $T$, and have no physical existence in the table). The bitmaps $B^i, 0 \leq i \leq s$ are called bit-slices, and their bit-values are defined this way:
$$
    C[j] = \sum_{i=0}^{s}{B^i[j] \cdot 2^i}
$$ where $C[j]$ is the $C$ value for the row with ordinal position $j$ in $T$, $B^i[j]$ is an indicator representing if $j$ is in bitmap $B^i$. In other words, $B^i[j] = 1$ if and only if bit $i$ in the binary representation of $C[j]$ is on. For a BSI denoted as $S$ representing the values of column $C$, we use $S[j]$ for representing the $C$ value for the row with ordinal position $j$. 

\begin{figure}
    \centering
    \includegraphics[width=1.0\linewidth]{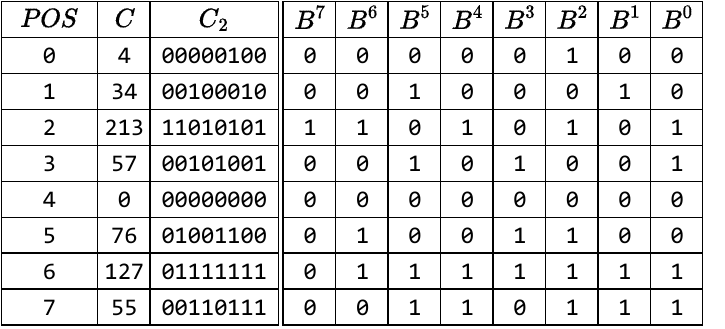}
    \caption{BSI Example}
    \label{fig:bsi-example}
\end{figure}

\autoref{fig:bsi-example} is an example of a BSI. In the figure, the column $C_2$ is the binary representation of column $C$. Each bit-slice of a BSI is like a vertical partition of a column. Range searches can be executed very efficiently using bit-sliced indexes. Some aggregate functions over values in a BSI, like the sum, average, median, and n-tile can also be executed efficiently on BSIs. Refer to \cite{o1997improved} for a complete analysis of aggregate functions and range searches. These algorithms make use of the bitmap operations (AND, OR, ANDNOT, XOR, etc.), described above, to operate on the BSI slices. 

\subsection{BSI Arithmetic}
\label{sec:bsi-arithmetic}

Bit-sliced index addition and other bit-sliced index operations like subtraction, multiplication, etc..., were defined in \cite{rinfret2001bit, rinfert2002term}. In these works, the focus was put on answering term matching (TM) queries. Later in some other works \cite{rinfret2008answering, guzun2015scalable}, BSI arithmetic was used to answer preference queries. Term matching queries work on the contents of textual documents, while preference queries work on combining different attributes of some relation. 

\begin{figure}
    \centering
    \includegraphics[width=0.8\linewidth]{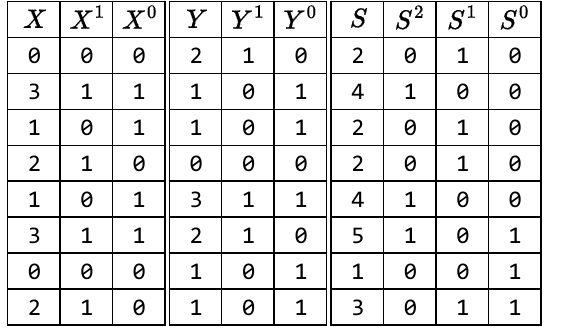}
    \caption{BSI Addition}
    \label{fig:bsi-add}
\end{figure}

Bit-sliced index addition for two BSIs $X$ and $Y$ works in the following way. We want to perform the addition $S = X + Y$ , where $X$ and $Y$ are the bit-sliced indexes for column $x$ and $y$, and $S[j] = X[j] + Y [j]$, for every row $r$ with ordinal position $j$. We perform the usual addition of binary numbers, not in a bit-by-bit manner, but by using bitmap operations described earlier. We first compute the low-order bit-slice of $S$ by doing $S^0 = X^0\ XOR\ Y^0$, and compute the carry bitmap with $C_0 = X^0\ AND\ Y^0$. Then we compute the next bit-slice of $S$ by doing $S^1 = X^1\ XOR\ Y^1\ XOR\ C_0$, and compute the new carry with $$C_1 = (X^1\ AND\ Y^1)\ OR\ [(X^1\ XOR\ Y^1)\ AND\ C_0].$$ We go on like this until we run out of slices. In the example of \autoref{fig:bsi-add}, both $X$ and $Y$ have two slices each, so we do $S^2 = C_1$. Remember that these bitmap operations are parallelized using the SIMD instructions.

The subtraction and multiplication of BSIs work in a similar way, mimicking digital logic algorithms \cite{reitwiesner1960binary} using bitmap operations. Furthermore, we can easily define comparison of BSIs using the same idea, which extend the capability of BSI arithmetic. As illustrated in \autoref{algo:bsi-lessthan}, \autoref{algo:bsi-equal} and \autoref{algo:bsi-notequal}, the comparison operator ($<, =, \neq,etc.$) takes two BSIs as input, and produces a binary-value BSI indicating the row-wise comparison results. We treat zero values in BSI as not existing in our scenarios, so we do not care the results of rows that have zero values in both BSIs when doing BSI arithmetic, and we set these results to zeros for better compression of the data with roaring bitmap.

It is easy to see that the computational complexity of addition, subtraction, comparison operators($<,=,\neq$), in-BSI aggregate functions and range searches grows linearly as the number of non-empty bit slices increases. While the complexity of general multiplication and division of BSI are $O(s_1s_2)$, $s_1$ and $s_2$ are the numbers of non-empty bit slices in each BSI, which seems to be slower. Fortunately, in the scenarios of this paper we only need the multiplication with one of the operators being binary, which makes the complexity also linear.

\begin{algorithm}[ht]
    \caption{Less Than Operator of BSIs}
    \label{algo:bsi-lessthan}
    \begin{algorithmic}[0]
    \STATE Given two BSI $X$ and $Y$ with $s$ bit slices, compute the less than comparison result $L$, which is a binary-value BSI (has only one bit slice) that satisfies $L[j] = 1$ if and only if $X[j] \neq 0, Y[j] \neq 0\ and\ X[j] < Y[j]$ for each row with ordinal position $j$.
    \STATE Initialize $L$ to empty BSI (with all value in it equal to 0)
    \FOR {$i=0$ to $s-1$}
        \STATE $L^0 \gets [(Y^i\ OR\ L^0)\ ANDNOT\ X^i]\ OR\ (Y^i\ AND\ L^0)$
    \ENDFOR
    \RETURN $L$
    \end{algorithmic}
\end{algorithm}

\begin{algorithm}[ht]
    \caption{Equal Operator of BSIs}
    \label{algo:bsi-equal}
    \begin{algorithmic}[0]
    \STATE Given two BSI $X$ and $Y$ with $s$ bit slices, compute the equal comparison result $E$, which is a binary-value BSI (has only one bit slice) that satisfies $E[j] = 1$ if and only if $X[j] \neq 0, Y[j] \neq 0\ and\ X[j] = Y[j]$ for each row with ordinal position $j$.
    \STATE Initialize $E$ to $X^0\ OR\ X^1\ OR\ ..\ OR\ X^{s-1}$
    \FOR {$i=0$ to $s-1$}
        \STATE $E^0 \gets E^0\ ANDNOT\ (X^i\ XOR\ Y^i)$
    \ENDFOR
    \RETURN $E$
    \end{algorithmic}
\end{algorithm}

\begin{algorithm}[ht]
    \caption{Not Equal Operator of BSIs}
    \label{algo:bsi-notequal}
    \begin{algorithmic}[0]
    \STATE Given two BSI $X$ and $Y$ with $s$ bit slices, compute the not equal comparison result $NE$, which is a binary-value BSI (has only one bit slice) that satisfies $NE[j] = 1$ if and only if $X[j] \neq 0, Y[j] \neq 0\ and\ X[j] \neq Y[j]$ for each row with ordinal position $j$.
    \STATE Initialize $NE$ to empty BSI (with all value in it equal to 0)
    \FOR {$i=0$ to $s-1$}
        \STATE $NE^0 \gets NE^0\ OR\ (X^i\ XOR\ Y^i)$
    \ENDFOR
    \RETURN $NE$
    \end{algorithmic}
\end{algorithm}

\section{Represent Experiment Data by BSI}
\label{sec:bsi-representation}

\subsection{Categories of Experiment Data in WeChat}

\begin{table*}[t]
  \caption{Categories of Experiment Data}
  \label{tab:data-categories}
  \begin{tabular}{ccl}
    \toprule
    Category & Object Described & Schema\\
    \midrule
    Expose Log & an experiment strategy& strategy-id, analysis-unit-id, randomization-unit-id, first-expose-date\\
    Metric Log & a metric at some date & date, metric-id, analysis-unit-id, value\\
    Dimension Log & an attribute at some date & date, dimension-name, analysis-unit-id, value\\
  \bottomrule
\end{tabular}
\end{table*}

\subsubsection{Expose Log}

Randomized traffic will be assigned to different strategies when an experiment is started. There are different randomization configurations for different scenarios. For example, an experimenter that wanted to test the latency of page views within strategy $A$ and $B$ may use page-view as the randomization unit for traffic allocation, and an experimenter that wanted to keep the consistent experience of WeChat users when the experiment is running should use user as the randomization unit, as thus the same user will be assigned to a fixed strategy in the experiment during the whole process.

In addition to the randomization unit, the analysis unit is another important concept, it is typically the denominator in a metric, e.g. page-view for page-click-rate and revenue-per-search, session for session-success-rate, etc. If we order different levels in a hierarchy \cite{deng2017trustworthy}, the randomization unit should always be higher or equal to the analysis unit (that is to say, the analysis unit is more fine-grained than the randomization unit when they are not the same unit). For example, any user-level metric would be ill-defined under page-level randomization, because the same user might be exposed to both the treatment and control strategies. In our scenarios, the analysis unit is equal to the randomization unit in most of the experiments. 
 
User requests with multiple unit ids will be exposed to the experiment strategies gradually when an experiment is started (e.g. a request of page-view issued by a user in a session will bring the page-view id, the session id and the user id in its context). The expose log of an experiment strategy consists of 4 columns: the strategy-id, the analysis-unit-id, the randomization-unit-id (e.g. related user-id of each page view) and the first-expose-date (the date when the experiment strategy starts to take effect on the unit), it is generated by aggregating raw log of exposed requests with same analysis-unit-id. We may generate multiple expose log for one experiment, because there may be metrics with different analysis units related to one experiment, for example, we may need to analyze page-view metrics and user-level metrics in a user randomized experiment.
 
\subsubsection{Metric Log}

Metric log describes metric values for each analysis unit in each day. For example, stay-time-per-user log is organized into 3 columns: date, user-id and stay-time, forwarding-count-per-session is organized into 3 columns: date, session-id and forwarding-count, active-days-per-user is organized into 3 columns: date, user-id and active-days. The metric log will be joined with the expose log by the analysis-unit-id for analyzing the metrics related to the experiment.  

\subsubsection{Dimension Log}

Dimension log describes some attributes of the analysis unit in each day. For example, user-age is a user-level dimension log that describes the age of each user, client-type is a page-view-level dimension log that describes the client used in a request. We also partition dimension log by date. Dimension log is used to filter specific analysis units out of expose log, making deep dive analysis of an experiment. 

It is worth mentioning that we generally have only one id (the analysis-unit-id) in metric log and dimension log, this is because that not the whole pipeline of generating the metric log and dimension log is inside our experiment platform, and the analysis-unit-id is the minimal requirement for matching related metric and dimension values to expose log when analyzing an experiment. With minimal requirement of the data, the pipeline of generating this data can be more flexible and maintainable. \autoref{tab:data-categories} summarizes the categories of experiment data.
 
\subsection{Segmentation}

The metric computation is based on the analysis unit. We segment the experiment data described above according to the analysis-unit-id. Specifically, we employ a hash function $HASH$ that is independent of traffic randomization process and calculate $segment\text{-}id = HASH(analysis\text{-}unit\text{-}id) \% 1024$. It is a deterministic randomization process for assigning all analysis units into different segments 0, 1, ..., 1023. Operations on each segment of data are identical when computing metrics, which makes segment the basic unit of parallel computing and load balancing.

\subsection{Bucketing and Statistical Inference}

A major challenge of metric computation is to perform statistical inference efficiently on the experiment data. As described earlier, not only a single value should be computed for an experiment metric, the variance of the metric (and the covariance between metrics) should be estimated correctly. 

We suggest that it is reasonable to assume the randomization units satisfy the stable unit treatment value assumption (SUTVA) \cite{rubin1980randomization, rubin1986comment, sekhon2008neyman}, which requires that "the observation of
potential outcome on one unit should not be affected by the particular assignment of treatments to the other units". It is an assumption about independence among randomization units, but goes beyond the concept of independence. If the assumption is not satisfied, for example, in a user-randomized experiment, a user who is exposed to experiment strategy $B$ will have different stay-time, depends on the strategy exposure of other users, then the stay-time computed by users in strategy $B$ with partial users in it can not stand for the real effect of strategy B when it is full launched, which makes the experiment conclusion invalid.

Based on the "independence" property of the randomization units, we introduce a bucketing procedure for estimating variance and covariance of metrics efficiently. Similar to what we did for segmentation, we introduce a deterministic randomization process for assigning all randomization units into different buckets 0, 1, ..., 1023, and we compute the metric value for each bucket separately. It can be seen as generating 1024 independent replicates of the metric values of an experiment strategy, and then bootstrapping \cite{johnson2001introduction, mooney1993bootstrapping} the variance and covariance of metrics. We proved that the bucket-based method of estimating the variance and covariance of metrics is correct in theory, and it makes the statistical inference procedures more efficient, check \cite{xiong2021covariance} for more details.
 
It is worth mentioning that the randomization unit and the analysis unit are the same in most of the experiments, in which case the segmentation and bucketing can be the same procedure.

\subsection{Position Encoding and BSI Representation}

\subsubsection{Position Encoding} 

\begin{figure}
    \centering
    \includegraphics[width=1\linewidth]{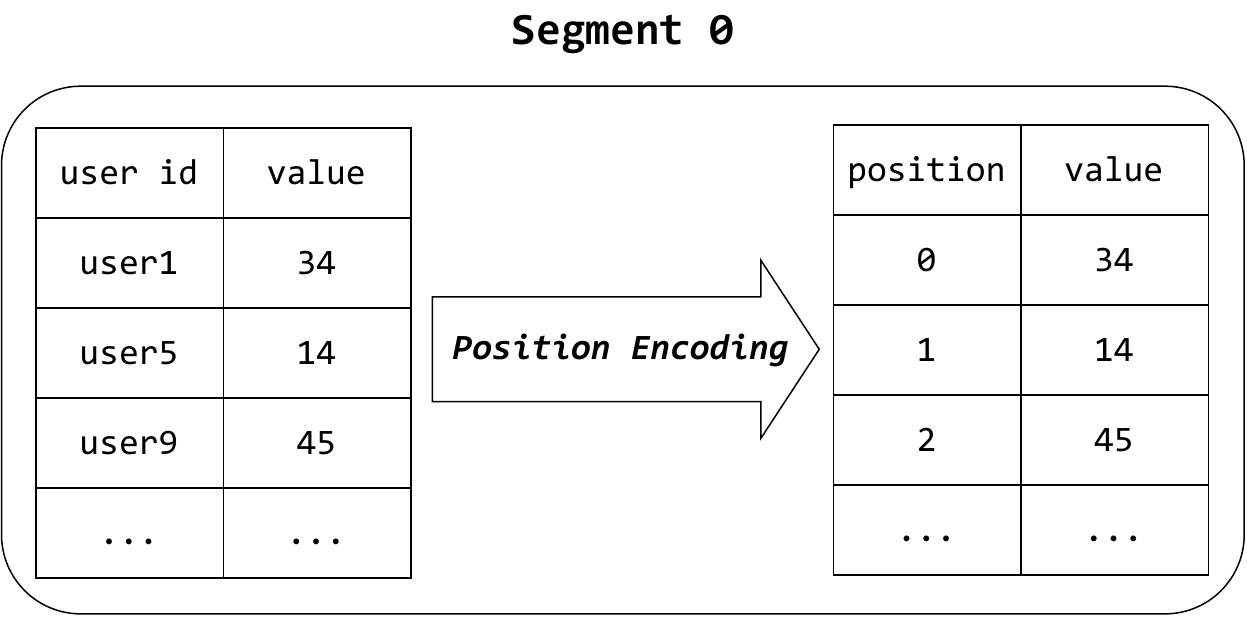}
    \caption{Position Encoding Example of Segment 0}
    \label{fig:position-encoding}
\end{figure}

We noticed that the analysis unit plays the center role when computing the metrics, and we introduce a position encoding process for making full use of BSI arithmetic's capability. Specifically, we encode each analysis-unit-id to a position within each segment independently, the position starts with 0 and increases sequentially for those ids that have not been encoded. As illustrated in \autoref{fig:position-encoding}, each user will be encoded to a fixed position, makes the value column be represented by BSI naturally and compactly (recall that zero values in BSI are treated not existing in our scenarios). Furthermore, we tend to encode the user-id with higher user engagement to smaller position as most as we can, to make the roaring bitmaps in BSI more compact and efficient.

\begin{table*}[t]
  \caption{BSI Representation of Experiment Data}
  \label{tab:BSI-rep}
  \begin{tabular}{ccl}
    \toprule
    Category & Object Described & BSI Representation\\
    \midrule
    Expose Log & an experiment strategy& segment-id, strategy-id, bucket-id(BSI), min-expose-date, offset(BSI)\\
    Metric Log & a metric at some date & segment-id, date, metric-id, value(BSI)\\
    Dimension Log & an attribute at some date & segment-id, date, dimension-name, value(BSI)\\
  \bottomrule
\end{tabular}
\end{table*}

\subsubsection{Expose Log Representation}

For each experiment strategy in a segment, there are 3 columns: analysis-unit-id, randomization-unit-id and first-expose-date in expose log. We first transform first-expose-date to 2 columns: min-expose-date and offset, because zeros in BSI are ignore, we start offset from 1, min-expose-date is a constant, representing the minimum date of original first-expose-date column. The offset column can be represented by BSI more efficiently than the original first-expose-date, because that offset has a smaller value range, which leads to less non-empty bitmaps in the BSI. Due to bucketing, the randomization-unit-id is not necessary to exist in exposed log, we use the bucket-id instead, finally, we get one constant: min-expose-date, and two BSIs: offset and bucket-id for representing the expose log for each experiment strategy, as illustrated in \autoref{tab:BSI-rep}.

\subsubsection{Metric and Dimension Log Representations}

For each metric (dimension) and each date in a segment, there are 2 columns: analysis-unit-id, value. We simply represent the value column by BSI after the position encoding process, as illustrated in \autoref{tab:BSI-rep}. 

\subsection{Efficiency of BSI Representation}
\label{sec:efficiency}

\begin{figure}
    \centering
    \includegraphics[width=1.0\linewidth]{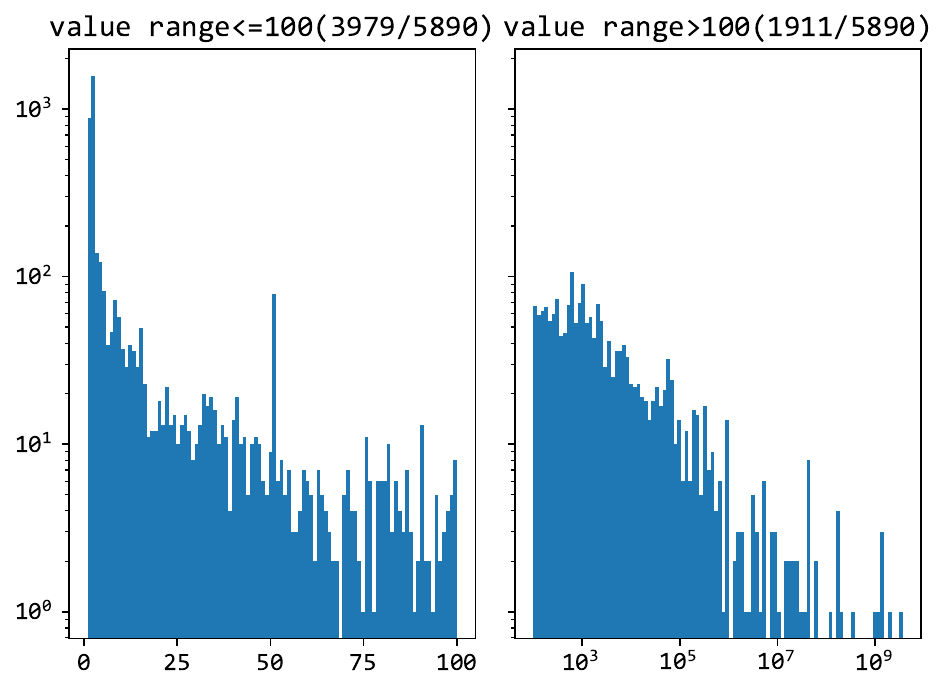}
    \caption{Value Range Cardinalities Distribution of 5890 Real World Metrics in One Day}
    \label{fig:value-range}
\end{figure}

\begin{figure}
    \centering
    \includegraphics[width=1.0\linewidth]{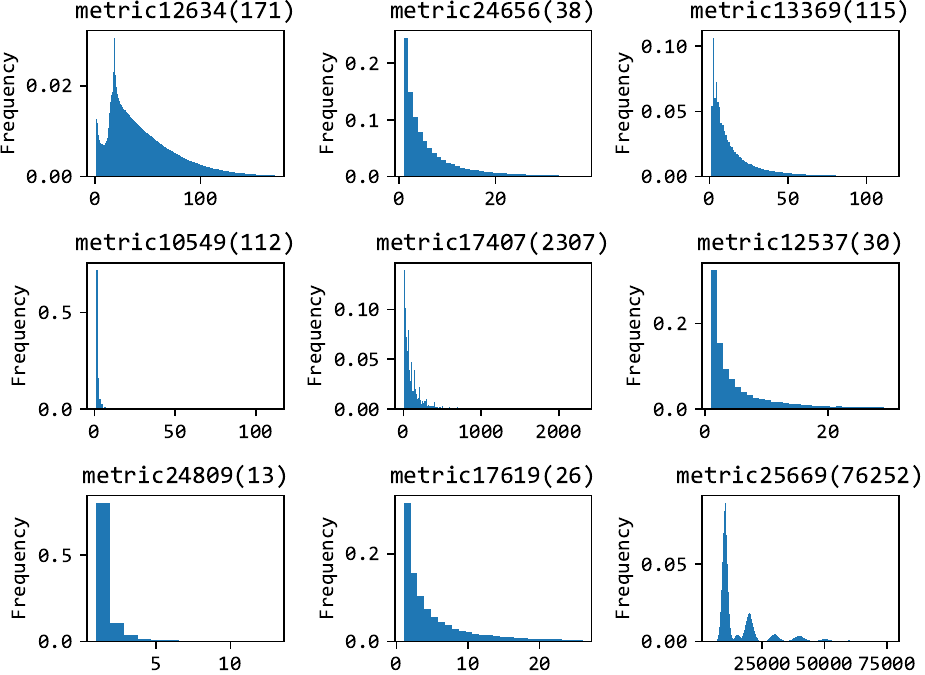}
    \caption{Metric Value Distribution Examples}
    \label{fig:value-dist}
\end{figure}

We use 5890 real world user-level metrics in our system for analyzing the characteristics of the data. As illustrated in \autoref{fig:value-range}, most of the metrics have a small value range in one day (for example, there are 3979 metrics out of 5890 have a value range with cardinality $\leq$ 100), and we also checked the distributions of the metric values, as illustrated in \autoref{fig:value-dist}, we found that the values are roughly distributed in ranges near to 0, following the Pareto principle \cite{dunford2014pareto}. We also observed the same characteristic in the expose data and the dimension data (for example, most of the users will be exposed to strategies in the beginning few days after the experiment has been started).

An important property of BSI with roaring bitmaps is that the data will be well compressed when the most of the values of BSI are concentrated in relatively small range and the position encoding is compact, because the roaring bitmap will compress the zeros of binary representation of each value and keep bitmap operations available on compressed data at the same time. This kind of compression is different from the data storage compression, it compresses the data size processed by CPU when executing the related operations on the data, for example, for raw data with two columns of size 1GB we need process 1GB data by CPU when adding these two columns, while for compressed BSI data we maybe only need process 100MB data by CPU, which improves the performance significantly if the addition operation on BSI are efficient enough compared to the addition operation on raw data.

We evaluate the performance of BSI representation in \autoref{sec:perf} with more details.

\section{Metric Computation by BSI Arithmetic}
\label{sec:metric-computation}

We now discuss the metric computation logic in our system. First we introduce some SQL operations (join, filter and aggregate) on the BSI representation, concrete metric computation scenarios are discussed later. 

\subsection{Join, Filter and Aggregate}
With the help of BSI arithmetic, we can perform join, filter and aggregate operations on experiment data represented by BSI. 

\subsubsection{Join}
All BSIs are naturally joined together by analysis-unit-id through the position encoding. We can access values of the same analysis unit in all BSIs by its encoded position.

\subsubsection{Filter}
The entire BSI can be filtered out with predicates on normal columns, for example, we can select the value BSI of a metric by metric-id and date:

\begin{verbatim}
SELECT value
FROM metric-log
WHERE metric-id = 8371 AND 
    date = '2024-02-27';
\end{verbatim}

Part of the values in BSI can be filtered out with predicates on BSIs, for example, we can select the expose information of a strategy whose analysis units are first exposed between the 2nd day and 5th day (the symbol $*$ is for multiplication of BSI, and the comparison is also performed on BSI):

\begin{verbatim}
SELECT min-expose-date, 
    bucket-id * (offset >= 2) * (offset <= 5), 
    offset * (offset >= 2) * (offset <= 5)
FROM expose-log
WHERE strategy-id = 8746325;
\end{verbatim}

\subsubsection{Aggregate}
Aggregate functions over values in a BSI, like the sum, average, median, n-tile can be executed efficiently, these function aggregate all values in a BSI into one numerical result, refer to \cite{o1997improved} for more details. Furthermore, we found that aggregate functions over BSIs, like sumBSI, maxBSI, mulBSI, distinctPos, can also be implemented efficiently by BSI arithmetic. These functions aggregate multiple BSIs into one BSI, for example, the sumBSI aggregate function add all BSIs together. The distinctPos aggregate function is a special aggregate for generating a binary BSI indicating all the distinct encoded positions where a value exists (i.e. there exists non-zero value on the position), it is used for calculating unique analysis units count (e.g. Unique Visitors) when computing metrics. The implementations of these functions are illustrated below.
\begin{verbatim}
sumBSI(X, Y) := X + Y;
maxBSI(X, Y) := X * (X > Y) + Y * (X <= Y)
mulBSI(X, Y) := X * Y
distictPos(X, Y) := (X > 0) OR (Y > 0)
\end{verbatim}

\subsection{Scorecard Computation}
\label{sec:scorecard}
The simplest form of experiment results is an experiment scorecard, it is a table consisting of a set of metrics and their movements in an experiment \cite{gupta2018anatomy}. It contains the observed metric values and the p-values that generated by the statistical inference (typically t-test) for each strategies in the experiment. At WeChat, it is common for an experiment scorecard to compute hundreds of metrics over tens of millions of end-users, and there are millions of strategy-metric pairs to be computed every day. 

Recall that the randomization unit and the analysis unit are the same in most of the experiments, in which case the segmentation and bucketing become the same procedure. For simplicity, we demonstrate the single-day scorecard computation for one strategy-metric pair in such case. The SQL below computes the bucket-value (by sum aggregate function) for each bucket, then the metric-value and related statistical inference are carried out based on these bucket-values.
\begin{verbatim}
SELECT t1.segment-id as bucket-id, 
    (t1.expose-date <= t2.date) as expose,
    (t2.value * expose) as filtered-value,
    sum(filtered-value) as bucket-value
FROM (
    SELECT segment-id, 
        (min-expose-date + offset - 1) as expose-date
    FROM expose-log
    WHERE strategy-id = 8764293
) as t1
INNER JOIN (
    SELECT segment-id, date, value
    FROM metric-log
    WHERE date = 'someday' AND
        metric-id = 8371
) as t2
ON t1.segment-id = t2.segment-id;
\end{verbatim}
For the case that the segment-id and the bucket-id are not the same, we need to sum the filtered-value by bucket-id, generating 1024 bucket-values for each segment, and then merge the bucket-value results of each segment. There may be some aggregate functions that cannot be merged through numerical bucket-values, like median (Non-decomposable aggregate functions \cite{jesus2014survey}), in such case we generate intermediate BSI format state for the bucket-value or employ some approximation algorithms. This is also the solution for merging non-decomposable aggregate function bucket-values of different dates, for example, we compute the $(value > 0)$ (a BSI format state $s$ representing unique visitors of a metric) as the bucket-value for a metric in different dates, and then use $sum(distinctPos(s))$ for merging these states into the unique-visitor count.

\subsection{Pre-Experiment Computation}
In \cite{deng2013improving} a technique was introduced that uses pre-experiment data to reduce the variance in experimentation metrics. We also implemented the method in our system. The computation of pre-experiment is similar to the scorecard computation, except that the expose-log is joined with $C$ successive days of the metric-log previous to the experiment start date. We also demonstrated the single-day pre-experiment computation for one strategy-metric pair just like done before.
\begin{verbatim}
SELECT t1.segment-id as bucket-id, 
    (t1.expose-date <= 'someday') as expose,
    (t2.value * expose) as filtered-value,
    sum(filtered-value) as bucket-value
FROM (
    SELECT segment-id, 
        (min-expose-date + offset - 1) as expose-date
    FROM expose-log
    WHERE strategy-id = 8764293
) as t1
INNER JOIN (
    SELECT segment-id, sumBSI(value) as value-sum
    FROM metric-log
    WHERE metric-id = 8371 AND date between 
        (expt_start_date - C) AND (expt_start_date - 1)
    GROUP BY segment-id     
) as t2
ON t1.segment-id = t2.segment-id;
\end{verbatim}

\begin{figure}
    \centering
    \includegraphics[width=1\linewidth]{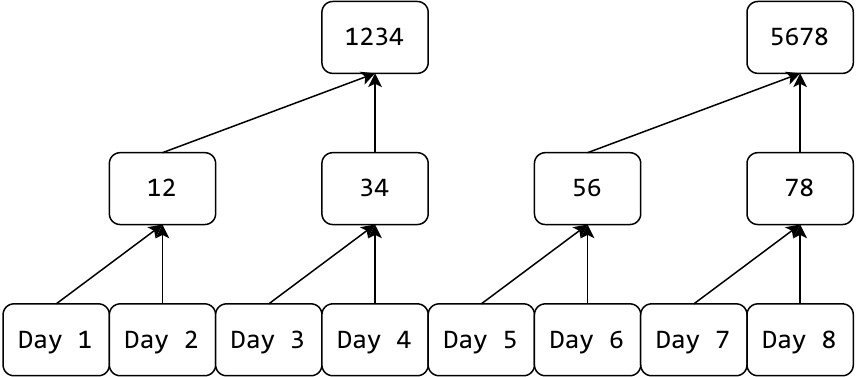}
    \caption{Pre-Aggregate Tree}
    \label{fig:tree}
\end{figure}

It is worth noting that $C$ successive days of metric-log are aggregated by $sumBSI$ first, this procedure can be accelerated by pre-aggregating the metric-log with a tree data structure. As illustrated in \autoref{fig:tree}, each non-leaf node of the tree is merged from its two children by some aggregate function over BSIs. We can compute aggregation of successive days more efficiently by merging less nodes in the tree, for example, performing $sumBSI$ for day 1 to day 7 can be accomplished by merging 3 nodes (1234, 56, 7) in the tree instead of 7 nodes.

\subsection{Deep Dive Analysis}
While the scorecard provides the overall strategy effect on a set of metrics, experiment owners should also have an ability to better understand metric movements. We provide an option to investigate the data by the analysis unit attributes (e.g. client-type) or time period (e.g. daily, weekly). Using this feature, experiment owners might discover heterogeneous effects on different attributes (e.g. a specific client-type might be causing a higher number of errors) and detect novelty effects (e.g. the treatment group that received a new feature engages with it in the first day and stops using it after that). 

This kind of analysis is too flexible to be pre-computed completely, and we compute them as needed by ad-hoc queries. The computation is also similar to the metric computation logic described earlier, but with an extra step for filtering the expose-log by dimension-log, the cost of this extra step is negligible because that we just need to analysis few strategy in deep dive analysis. We demonstrate the expose filtering example below, which filters the expose-date of the analysis units with client-type = 1 and client-version > 134.
\begin{verbatim}
SELECT t1.segment-id as segment-id, 
    (t1.expose-date * t2.dim-filter) as expose-date
FROM (
    SELECT segment-id, 
        (min-expose-date + offset - 1) as expose-date
    FROM expose-log
    WHERE strategy-id in (8764293,8764294,8764295)
) as t1
INNER JOIN (
    SELECT segment-id, mulBSI(filter) as dim-filter
    FROM (
        SELECT segment-id, (value = 1) as filter
        FROM dimension-log
        WHERE dimension-name = 'client-type' AND
            date = 'someday'
        UNION ALL
        SELECT segment-id, (value > 134) as filter
        FROM dimension-log
        WHERE dimension-name = 'client-version' AND
            date = 'someday'
    )
    GROUP BY segment-id
) as t2
ON t1.segment-id = t2.segment-id
\end{verbatim}

\section{System Architecture}
\label{sec:arch}

\begin{figure*}
    \centering
    \includegraphics[width=1\linewidth]{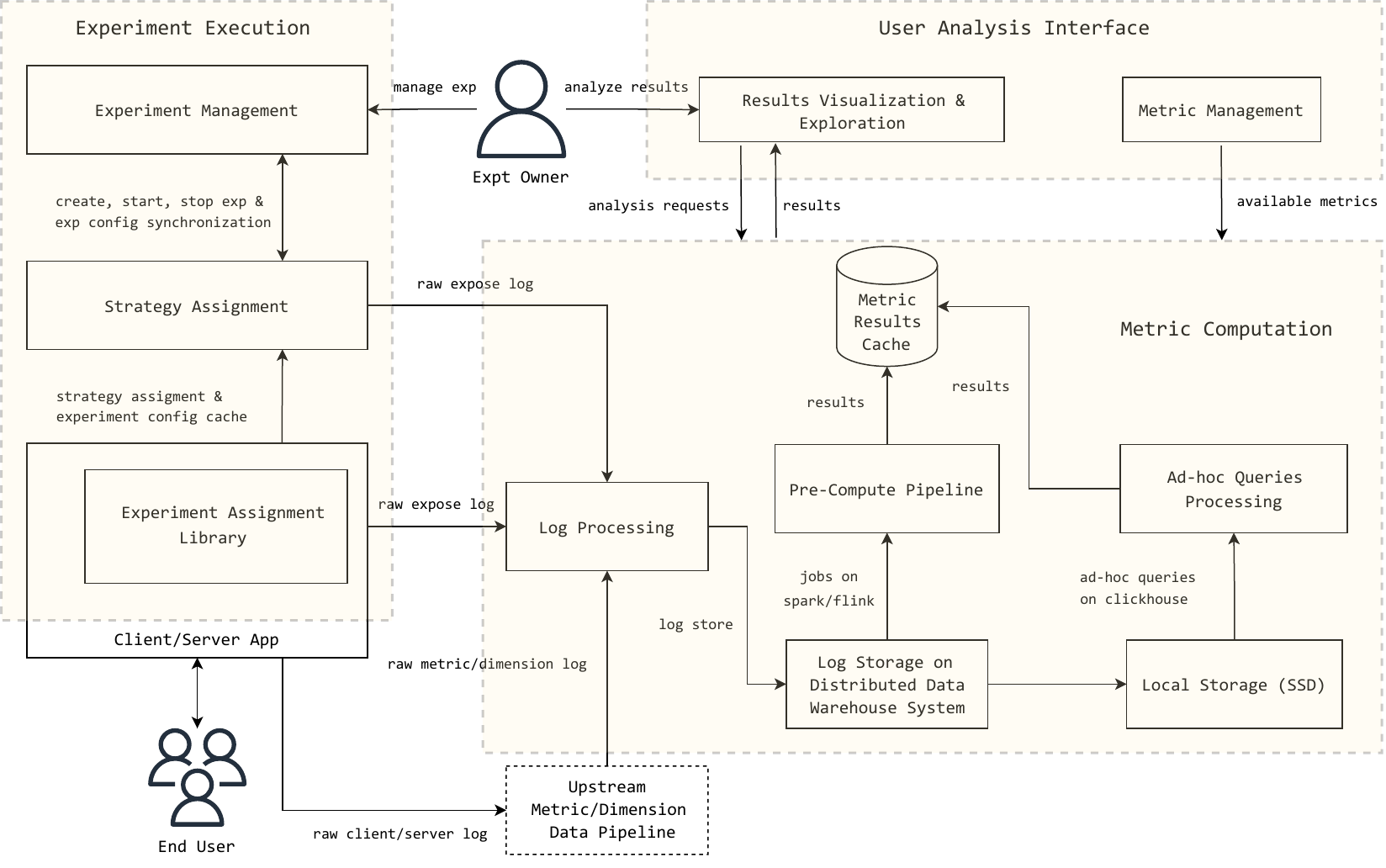}
    \caption{Architecture for the Metric Computation}
    \label{fig:arch}
\end{figure*}

\subsection{Overall Architecture}
The architecture of the metric computation is illustrated in \autoref{fig:arch} in the next page. The raw expose log and the raw metric/dimension log are first processed, then converted to BSI representations and stored on a distributed data warehouse system. Routine computations like the scorecard computation and the pre-experiment computation are executed by pre-compute pipeline using Apache Spark \cite{zaharia2010spark, zaharia2012resilient}. Ad-hoc queries like deep dive analysis are processed by Clickhouse \cite{imasheva2020practice} (an open-source columnar database management system for online analytical processing). 

\subsection{Pre-Computation by Spark}
We submit jobs to spark clusters for pre-computing the metrics every day. The experiment data in BSI format is read from the distributed data warehouse system through network by spark jobs. Each job computes a batch of strategy-metric pairs for better utilizing network traffic. The results are cached for user analysis later in the day. The BSI related operations are implemented in the spark framework. Recall that the BSI operations are based on roaring bitmap operations, which can be parallelized by exploiting SIMD instructions, we delegate these bitmap operations to SIMD implementations through Java Native Interface \cite{liang1999java}.

\subsection{Ad-hoc Queries by Clickhouse}
\label{sec:ch-arch}

\begin{figure}[h!]
    \centering
    \includegraphics[width=1.0\linewidth]{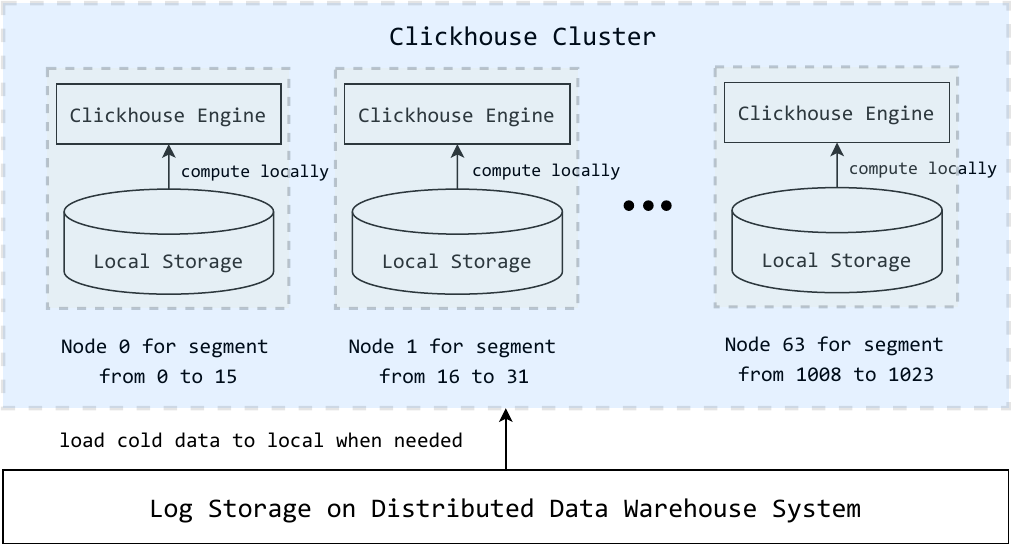}
    \caption{Ad-hoc Queries Processing}
    \label{fig:ch}
\end{figure}

The critical issue of add-hoc queries is answering them as quickly as possible, thus we introduce the Clickhouse with fast local storage (typically local SSD) in our system. The details of ad-hoc queries processing are illustrated in \autoref{fig:ch}, generally each segment of data will be located in one node of the Clickhouse cluster, and the BSI related operations are implemented in the Clickhouse engine for processing the queries. Queries are processed in parallel and locally in each node, makes the latency as low as possible. We only keep the hot data (e.g. data with recent date or visited recently) for reducing the storage cost of local storage, the cold data located in the distributed data warehouse system will be loaded to local storage when needed.


\section{Performance Evaluation}
\label{sec:perf}
In this section we evaluate the performance of BSI representation and the real world computation performance of our system.

\subsection{BSI Representation Evaluation}
In \autoref{sec:efficiency} We use 5890 real world user-level metrics in our system for analyzing the characteristics of the data, and conclude that we can compress the data through the BSI representation and keep the operations on it available without decompression. In this subsection we evaluation the storage and computational efficiency of the BSI representation, or BSI format.

\subsubsection{Storage Evaluation}
 We choose 105 user-level core metrics, these core metrics are the most important metrics of one of the vital business scenarios in WeChat. The value range cardinalities distribution of the these metrics in one day are illustrated in \autoref{tab:value-range-105}, the table shows that these metrics tend to have bigger value ranges compared to the 5890 metrics, and they should be enough for evaluation usage (the performance will be better if the evaluation is carried out with the 5890 metrics). We do not use more metrics here, because too many metrics will lead to too much storage and computational cost.

\begin{table}[t]
  \caption{Value Range Cardinalities Distribution of the 105 Core Metrics}
  \label{tab:value-range-105}
  \begin{tabular}{ccc}
    \toprule
    Range Card (One Day)& Number of Metrics & Proportion\\
    \midrule
    (0, 10] & 33& 31.4\% \\
    (10, 100] & 4& 3.8\% \\
    ($10^2$, $10^3$] & 26& 24.8\% \\
    ($10^3$, $10^4$] & 18& 17.1\% \\
    ($10^4$, $10^5$] & 12& 11.4\% \\
    ($10^5$, $10^6$] & 5& 4.8\% \\
    ($10^6$, $10^7$] & 5& 4.8\% \\
    ($10^7$, $10^8$] & 2& 1.9\% \\
  \bottomrule
\end{tabular}
\end{table}


\begin{table}[t]
  \caption{Storage of 105 Core Metrics in a Month(29 days)}
  \label{tab:BSI-storage}
  \begin{tabular}{cccc}
    \toprule
    Format & Rows & Compressed Size(LZ4) & Original Size\\
    \midrule
    Normal & 890 billion& 4.1 TB & 15.6 TB\\
    BSI & 3.1 million & 1.6 TB & 1.7 TB\\
  \bottomrule
\end{tabular}
\end{table}

Then we generate the metric-log within a month using a normal format, the schema of the normal format is (segment-id UInt16, date UInt32, metric-id UInt32, user-id UInt32, value UInt32), and the schema of the BSI format is (segment-id UInt16, date UInt32, metric-id UInt32, value BSI). We compared the storage cost of these two formats, the results are illustrated in \autoref{tab:BSI-storage}. The compressed size of the BSI representation is about half of the size of the normal representation, and it is worth noting that the original size of BSI representation is roughly the same as the compressed one, because that the BSI representation is already a compressed format as described earlier. 

\begin{table}[t]
  \caption{Details of Three Typical Metrics in One Day}
  \label{tab:typical-3-metrics}
  \begin{tabular}{ccccc}
    \toprule
    Metric & Rows & Normal Size & BSI Size & Value Range\\
    \midrule
    A & 316 million& 2.9 GB & 140 MB & (0, 1]\\
    B & 34 million&  324 MB & 86 MB & (0, 50]\\
    C & 510 million& 4.7 GB & 2.0 GB & (0, 21600]\\
  \bottomrule
\end{tabular}
\end{table}

\subsubsection{Computational Evaluation}
Next we evaluate the computational cost of the BSI representation. We choose 3 typical metrics with different value ranges and rows. Raw data of each metric in one day is represented as rows of (segment-id, user-id, metric-value) with normal format, and there is at most one row for each user in one day (that is, aggregated by user-id within one day). Raw data of the metrics is converted to BSI representation for evaluation. The details of the 3 metrics in one day are illustrated in \autoref{tab:typical-3-metrics}, The size of data that processed by CPU with normal format (Normal Size column) is only affected by number of rows, while the size of BSI is determined by number of rows and value range (value distribution). Metric $C$ has a big value range and the maximum number of rows, thus has a maximum BSI size in one day, metric $B$ has a bigger value range than metric $A$, but also has much less rows than metric $A$, and the final BSI size of metric $B$ is smaller than metric $A$. 

For each metric, we use one CPU core for calculating the sum of metric values for each user in two days (that is, within each segment, performing $sumBSI$ for two BSIs or aggregating two days raw data by user-id). The evaluation program is single threaded and written in C++, compiled with $-O3$ optimization option, and runs 10 times repeatedly on a machine with a 2.4GHZ CPU. The performance results are listed in \autoref{tab:time-3-metrics}, it shows that the computation of BSI representation is much more efficient than that with the normal format, the main reason is that the size of data processed by CPU with BSI representation is much more smaller than that with normal format, and the SIMD instructions also play an important role in performance improvement.

We conclude that the performance gain of the BSI computation mainly comes from the compression of data when processed by CPU and the SIMD instructions when performing BSI operations.

\begin{table}[t]
  \caption{Average Time of Computation with Normal Format and BSI Format}
  \label{tab:time-3-metrics}
  \begin{tabular}{cccc}
    \toprule
    Format & Metric A & Metric B & Metric C \\ 
    \midrule
    Normal & 59.2 seconds& 7.3 seconds & 94.3 seconds\\
    BSI (no SIMD) & 1.2 seconds& 3.8 seconds & 18.3 seconds\\
    BSI (AVX2) & 0.6 seconds& 1.3 seconds & 10.5 seconds\\
  \bottomrule
\end{tabular}
\end{table}

\subsubsection{Extra Cost in Log Processing}
We evaluated the extra computational cost in log processing derived from conversion to the BSI format in the same settings. For each metric, we converted a day's worth of data from the normal format to the BSI format using a straightforward method, extracting bits from each metric value and setting these bits into corresponding bitmaps. We also observed that the conversion process can be more efficient when the data is pre-sorted by user-id in the normal format. By splitting the pre-sorted rows into blocks and processing each block with neighbouring user-ids, we set bits extracted from metric values in the block into adjacent containers in the roaring bitmap. This approach achieves better cache locality, enhancing the overall efficiency of the conversion. The results are illustrated in \autoref{tab:log-process-time-3-metrics}. It is easy to see that the conversion cost is mainly related to the number of rows and the value range of the metric, more rows and bigger value range cost more CPU time. In our situation, most metrics have relatively small value ranges, and the total computation cost (conversion and BSI computation) is still much lower than that with normal format in our scenarios. 

Furthermore, we have monitored the overall cost (CPU hours and latency) of log processing in our real system, and the results show that the log processing cost remains almost the same as before, the reason is that there are more IO-bound operations in log processing, so the BSI conversion cost does not become a bottleneck in this procedure.

\begin{figure}
    \centering
    \includegraphics[width=1\linewidth]{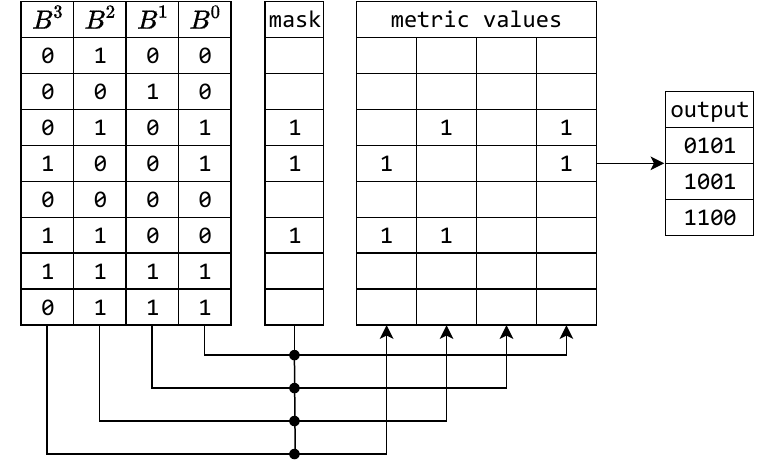}
    \caption{Convert back to Normal Format within a Container}
    \label{fig:convert-back}
\end{figure}

\subsubsection{Cost of Converting Back to the Normal Format}
\label{sec:conversion-cost}
We also evaluated the cost of converting data back to the normal format in the settings above. For each metric, we converted a day's worth of data from the BSI format to the normal format using both a straightforward method and a per-bitmap method. Using the straightforward method, we collect the bits of the metric-value for a user from each bitmap and combine them into a single value. The problem of the straightforward method is that we need to check all the bitmaps for each metric value, even if there is only one bit in the value. Additionally, the scattered bitmap access reduces the cache hit rate. To resolve the problem, we extract the bits in a per-bitmap manner, as illustrated in \autoref{fig:convert-back}, for each bitmap, bits identified by a mask (indicating the user-ids of interest) within a container are first extracted and then set into corresponding positions of metric values, the output is then generated by retrieving the metric values corresponding to the mask, this process is repeated for each container of the roaring bitmap. This approach avoids the extra cost of processing zero bits in metric values and achieves better cache locality. Additionally, the bit extraction process can also be accelerated by SIMD instructions when necessary. The evaluation results, illustrated in \autoref{tab:to-column-time-3-metrics}, shows that the conversion back process is very efficient with the per-bitmap method. Furthermore, the output of the conversion is naturally sorted by user-id within the mask, and can be efficiently converted to the BSI format again.

\begin{table}[t]
  \caption{Average Time of Converting to BSI format}
  \label{tab:log-process-time-3-metrics}
  \begin{tabular}{cccc}
    \toprule
    Method & Metric A & Metric B & Metric C \\ 
    \midrule
    Straightforward &4.2 seconds& 0.9 seconds & 34.9 seconds\\
    Pre-sorted & 3.9 seconds & 0.8 seconds & 12.9 seconds\\
  \bottomrule
\end{tabular}
\end{table}

\begin{table}[t]
  \caption{Average Time of Converting back to Normal Format}
  \label{tab:to-column-time-3-metrics}
  \begin{tabular}{cccc}
    \toprule
    Method & Metric A & Metric B & Metric C \\ 
    \midrule
    Straightforward &44.6 seconds& 10.0 seconds & 164.6 seconds\\
    Per-bitmap & 2.0 seconds & 1.1 seconds & 8.7 seconds\\
  \bottomrule
\end{tabular}
\end{table}

\subsection{Pre-Computation Evaluation}
Pre-computation performance is also evaluated on the 105 core metrics. There are about 240,000 strategy-metric pairs to be computed in November 21, 2023 in our system, these pairs involves about 8,500 strategies and each strategy contains an average of 21 million exposed users. We evaluate the CPU hours consumed for computing the scorecard results of these pairs using two methods. The BSI-based method is described in \autoref{sec:scorecard}, while the normal representation based method is described as follow (this is also the method used in our system before the deployment of the BSI method): 
\begin{itemize}
    \item Represent the expose-log by the normal format: (segment-id UInt16, strategy-id UInt32, bucket-id UInt16, first-expose-date UInt32) 
    \item Implement the scorecard computation logic with Spark SQL \cite{armbrust2015spark} on expose-log and metric-log with the normal format 
    \item Spark SQL jobs are executed on our clusters, we split the strategy-metric pairs into several jobs so that the total number of cores of the job will not be too large (no more than 2000 cores)
\end{itemize}

The results are illustrated in \autoref{tab:cpu-hours}, the CPU hours of the BSI-based method are about a quarter of that with the normal format. 

\begin{table}[t]
  \caption{CPU Hours for Pre-computation for 105 Metrics}
  \label{tab:cpu-hours}
  \begin{tabular}{ccc}
    \toprule
    Format of Representation &  CPU Hours Consumed\\
    \midrule
    Normal & 22712\\
    BSI & 5446\\
  \bottomrule
\end{tabular}
\end{table}

\subsection{Ad-hoc Queries Evaluation}
The latency of the ad-hoc queries are evaluated on the 105 core metrics. We choose an experiment containing 3 strategies with an average of 200 million exposed users (a huge experiment!), and issue ad-hoc queries for computing the 105 core metric results within a specific week. These ad-hoc queries will be processed on the Clickhouse cluster described in \autoref{sec:ch-arch} by the BSI-based method and the normal representation method (repeat 10 times). The normal representation based method is similar to the BSI-based method, and is described as follow  (this is also the method used in our system before the deployment of the BSI method):
\begin{itemize}
    \item For each segment, use a bitmap for representing the exposed users in each day in this week, and cache these bitmaps in memory (join is slow in Clickhouse, we do not join expose-log and metric-log with the normal format here, we use bitmaps for expose-log instead)
    \item Scan the metric-log with the normal format, and filter the rows which satisfy expose condition: the user-id of the row is contained in the expose bitmap
    \item Aggregate the metric values of each segment on each Clickhouse node in parallel
\end{itemize}

The results are illustrated in \autoref{tab:ch-latency}, the latency of the BSI-based method is reduced significantly compared to that with the normal format. 

\begin{table}[t]
  \caption{Latency for Ad-hoc Queries on 105 Metrics}
  \label{tab:ch-latency}
  \begin{tabular}{ccc}
    \toprule
    Format of Representation &  Average Latency\\
    \midrule
    Normal & 22.3 seconds\\
    BSI & 6.0 seconds\\
  \bottomrule
\end{tabular}
\end{table}

\section{Expressive Power and Limitation}
\label{sec:limitation}

BSIs can be considered as unsigned numerical vectors that support element-wise elementary arithmetic between vectors ($+$, $-$, $\times$, $\div$, $<$, $=$, etc) and aggregate operators over values in the vector (sum, max, min, etc), the proposed BSI-based method has the power of expressing algorithms that can be implemented by these operators, for example, RSME can be computed with this method by:
\begin{equation}
\begin{split}
&RMSE(v)^2=\frac{sum(v_i^2)}{n}-(\frac{sum(v_i)}{n})^2\\
&=\frac{sum(mulBSI(v, v))}{sum(gtBSI(v, 0))}-(\frac{sum(v)}{sum(gtBSI(v, 0))})^2
\end{split}
\end{equation} 
where $v$ is a vector containing metric values of each user, $v_i$ is the value at position $i$ of the vector (zero value for nonexistence), $n$ is the number of non-zero values in $v$, $mulBSI$ and $gtBSI$ mean $\times$ and $>$ for BSIs. 

For other algorithms that are difficult to implemented by BSI operators, we can convert the BSI format back to normal format on the fly for adaptation, as illustrated in \autoref{sec:conversion-cost}, the conversion process is efficient and won't be bottleneck of the pipeline in most cases. 

The main limitation of the method is the performance degradation of BSI operations under certain circumstances. Especially for multiplication and division, as described in \autoref{sec:bsi-arithmetic}, unlike other linear complexity operators, the time complexity of $mulBSI$ and $divBSI$ are $O(s1s2)$, the performance gain of these non-linear complexity operators will diminish rapidly in certain cases. For example, when values in BSI are distributed across large ranges or are uniformly distributed, the BSI representation cannot compress the data well. In such cases, we can convert the BSI format back, perform the multiplication (or division), and then convert it back to BSI to avoid severe performance degradation. However, in our situations, values always follow the Pareto principle as described in \autoref{sec:efficiency}, and we do not need general multiplication and division in most cases either, the limitation described above won't be a major concern.

\section{Conclusion}
\label{sec:conclusion}

We implemented a large-scale metric computing system using the BSI approach. We started by analyzing the characteristics of the data in our situation. Our observations lead to a BSI representation of the experiment data. The BSI representation with roaring bitmaps is an already compressed structure, and the arithmetic operations on it are performed at the compressed data directly, thus being very efficient. We implemented the metric computation logic in our system using BSI operations, and evaluated the storage and computational efficiency of the BSI representation, and also evaluated the real-world performance in the pre-computation and ad-hoc computation scenarios. 

With the deployment of the new system, we save a lot of computing resources, and support faster interactive explorations of the experiment data. The system has successfully met the analysis needs of experiment owners, and is widely used within WeChat as an important data-driven tool for decision making. It allows us to continuously improve the user experience of one of the world's largest social media platforms.

\begin{acks}
We sincerely thank the VLDB reviewers for their insightful suggestions, which have greatly improved this work. We would like to express our deepest gratitude to Xu Dong, Wencheng Liu, Hongyi Sun, and Lv Feng for their support in optimizing the Clickhouse engine. Their expertise has been instrumental in enhancing our system's performance. Our appreciation extends to Penglei Zhao for his contributions to implementing BSI arithmetic in Clickhouse, and to Long Wang for his work on BSI arithmetic in Spark. Special thanks go to Qi Sheng and Jiangtao Zhang for their assistance with implementing ad-hoc queries in Clickhouse. We are also deeply grateful to Senlie Zheng and Yichen Liu for their support in migrating from the old system to the new one. Their guidance was vital for a smooth transition.
\end{acks}


\bibliographystyle{ACM-Reference-Format}
\bibliography{sample}

\end{document}